\documentclass[aps, floatfix, twocolumn, a4paper, showpacs, nofootinbib,
superscriptaddress, 10pt]{revtex4}
\usepackage{graphicx,float,tikz}
\usepackage[all]{xy}
\newcommand\ringring[1]{%
  {
   \mathop{\kern0pt #1}\limits^{
     \vbox to-1.85ex{
       \kern-2ex 
       \hbox to 0pt{\hss\normalfont\kern.1em \r{}\kern-.45em \r{}\hss}%
       \vss 
     }
   }
  }
}
\usepackage{amsmath,upgreek}
\usepackage{amssymb}
\usepackage{color}
\usepackage{epsfig}		
\usepackage{graphicx,epstopdf}
\usepackage{subfigure}
\usepackage{pdfpages}
\usepackage[colorlinks,hyperindex]{hyperref}

\definecolor{green1}{RGB}{0,128,0} 
\hypersetup{backref=true,pagebackref=true}
\hypersetup{%
  colorlinks = true,
  linkcolor  = blue,
  citecolor = cyan,
}
\usepackage{bookmark,textgreek}
\usepackage{hyperref,color,xcolor}
\hypersetup{hidelinks,hyperindex=true,colorlinks=true,breaklinks=true,urlcolor= blue}
\hypersetup{%
  colorlinks = true,
  linkcolor  = blue
}

\usepackage[normalem]{ulem} 

\newcommand\orcidroldao{{\href{https://orcid.org/0000-0003-3978-532X}{\orcidicon}}}
\newcommand{\orcidicon}{%
	\begin{tikzpicture}
	\draw[lime, fill=lime] (0,0)
		circle [radius=0.16]
		node[white] {{\fontfamily{qag}\selectfont \tiny ID}};
	\draw[white, fill=white] (-0.0625,0.095)
		circle [radius=0.007];
	\end{tikzpicture}	\hspace{-2mm}
}

\newcommand{\bes}{\begin{subequations}}
\newcommand{\ees}{\end{subequations}}
\def\beq{\begin{eqnarray}}
 \newcommand{\blt}{\textcolor{black}}
  
\def\eeq{\end{eqnarray}}
\def\be{\begin{equation}}
\def\ee{\end{equation}}

\begin{document}

\title{Odd-spin glueballs, AdS/QCD and information entropy}

\author{D. Marinho Rodrigues}
\email{diegomhrod@gmail.com}
\affiliation{Federal University of ABC, Center of Mathematics,  Santo Andr\'e, 09580-210, Brazil}
\affiliation{Federal University of ABC, Center of Physics,  Santo Andr\'e, 09580-210, Brazil}
\author{R. da Rocha\orcidroldao\!\!}
\email{roldao.rocha@ufabc.edu.br}
\affiliation{Federal University of ABC, Center of Mathematics, Santo Andr\'e, 09580-210, Brazil}
\affiliation{}

\begin{abstract}
Odd-spin glueballs in the dynamical  AdS/QCD model are scrutinized, in the paradigm of the configurational entropy (CE). Configurational-entropic Regge trajectories, that relate the CE underlying odd-spin glueballs to their mass spectra and spin, are then engendered. They predict the mass spectra of odd-spin glueballs, besides pointing towards the configurational stability of odd-spin glueball resonances.  The  exponential modified dilaton with logarithmic anomalous dimensions comprises the most suitable choice to derive the mass spectra of odd-spin glueballs, compatible to lattice QCD. It is then used in a hybrid paradigm that takes both lattice QCD and the AdS/QCD correspondence into account.  
\end{abstract}
\maketitle

\section{Introduction}

The configurational entropy (CE) consists of the part of the entropy of a system that is related to representative correlations among its constituents. The CE measures the rate in which information can be compressed into any given system, in the  lossless regime where one can recover the entire original message by decompression. Any compressed message presents the same amount of information as the original one, however,   communicated along with fewer characters, being thus less redundant, complying with Shannon's \blt{information} theory \cite{Shannon:1948zz}.  \blt{In his seminal work, Shannon changed information, from an unestablished  concept, into a precise theory that has underlain the digital revolution. It has inspired the development of all modern data-compression algorithms and error-correcting codes as well. The CE can be interpreted as the measurement of the shape complexity and the degree of coherent order in a system. It can be applied to a considerable variety of  problems when one introduces some distribution portraying the system that is related to physical quantities under scrutiny.
To determine the informational features of physical systems, the  probability distribution is used in the momentum space, whereas different wavelengths impart the generation of correlations across the system \cite{Gleiser:2011di,Gleiser:2018kbq,Bernardini:2016hvx,Gleiser:2012tu}. }

The CE can be calculated for any physical system, once a spatially-localized, Lebesgue-integrable, scalar field that describes the system is chosen \cite{Sowinski:2015cfa}. \blt{The CE interplays dynamical and informational contents of any physical system that has localized energy configurations \cite{Gleiser:2012tu}.
This approach was shown to be fruitful to study solitons,  bounces, and critical bubbles, also including phase transitions \cite{Sowinski:2017hdw}. The CE was also employed to distinguish dynamical systems among regular, entirely random, and also chaotic evolution in QCD systems, providing observables and techniques
to comprehend  physical phenomena \cite{Ma:2018wtw}.}

\blt{In the CE approach, describing a physical system requires a spatially-localized scalar field, usually taken as the energy density, namely the $T_{00}$ component of the stress-energy-momentum tensor.} Nevertheless, scattering amplitudes and cross-sections are also successful pickings for representing the localized scalar field in QCD \cite{Karapetyan:2018oye,Karapetyan:2018yhm}. 
The CE is a precise tool for investigating and predicting relevant features of elementary particles and their resonances in AdS/QCD \cite{Bernardini:2016hvx,Bernardini:2018uuy,Ferreira:2019inu}, supported by experimental data in the Particle Data Group (PDG) \cite{pdg1}. Besides, the CE has made important and precise predictions for current runnings in several experiments.  
\blt{In the past five years, the CE has been successfully shown to be a prominent paradigm to examine the essential features of the  AdS/QCD correspondence. In particular, the CE is a very suitable instrument for investigating QCD. 
Several light-flavor mesons, tensor mesons, baryons, scalar glueballs, quarkonia, and pomerons were scrutinized using the CE in AdS/QCD  \cite{Bernardini:2016hvx,Bernardini:2018uuy,Ferreira:2019inu,Bernardini:2016qit,Ferreira:2019nkz,Colangelo:2018mrt,Ferreira:2020iry,MarinhoRodrigues:2020ssq,Braga:2017fsb,Braga:2018fyc,Braga:2020myi,Braga:2020hhs}. Other important applications in several physical contexts were also accomplished \cite{Lee:2018zmp,Correa:2016pgr,Cruz:2019kwh,Bazeia:2018uyg,Alves:2014ksa,Alves:2017ljt}.}

The AdS/QCD correspondence consists of a successful scheme to report QCD as a theory that is dual to gravity in a codimension-one bulk. It emulates AdS/CFT in a special circumstance where QFT does not exactly represent a conformal field theory \cite{Maldacena:1997re,Witten:1998qj,gub}. 
Since the bulk AdS space is invariant under conformal transformations, the associated string theory consists of a gauge theory that is conformally invariant, with features resembling  QCD, in the  large-$N_c$ regime. For establishing QCD from a dual gravity setup, conformal symmetry has to be broken on the AdS boundary. 
This is justified by the fact that QCD is approximately conformal when the high-energy regime sets in. 
Introducing either a soft wall or a hard wall is a procedure that  adjusts the AdS bulk. Each one of these choices is  suitable for different purposes, yielding precise phenomenological  aspects \cite{BoschiFilho:2002ta,BoschiFilho:2002vd}. The broken conformal symmetry additionally emulates confinement  
\cite{EKSS2005,Karch:2006pv,Brodsky:2014yha}. 

\blt{QCD conjectures the existence of bound states of gluons, therefore encompassing glueball states \cite{Gutsche:2016wix}, and also the odderon, consisting of a 
$C=-1=P$ counterpart of the $C=1=P$ pomeron.  
Current works predict that the exchange of the odderon yields a phenomenological deviation between $pp$ and $p\bar{p}$ high energy scattering. Besides, the central diffractive production of $f_0(980)$ and $f_2(1270)$ mesons at LHC has the odderon playing a relevant role in several production channels \cite{Machado:2011vh}. The 
photoproduction of $\pi^0$, $f_2(1270)$, and
$a_2(1320)$ mesons allows a clear experimental signature of odderons.}
\blt{Experiments showed the existence of the odderon in perturbative QCD \cite{adlo}. The maximal odderon, introduced by Heisenberg \cite{6}, occupies a prominent role in QCD, 
yielding reasonable deviations between particle-antiparticle and particle-particle cross-sections
\cite{bart,kwie}, with experimental signatures \cite{Martynov:2017zjz,Khoze:2018bus}. 
Currently, the odderon is a pivotal tool in QCD, being studied in  experiments at ATLAS -- LHC and RHIC as well, when measuring hadronic  scattering amplitudes \cite{Avila:2006wy}. 
One can, therefore, assert that deeply studying the odderon is
a current prominent test of QCD. }

The main goal in this work consists of employing the well-established   CE approach, widely used in QCD, to study odd-spin glueballs  in the AdS/QCD setup.  CE Regge trajectories represent the leading apparatuses to derive odd-spin glueballs mass spectra, in a hybrid model that employs both lattice QCD and the AdS/QCD correspondence.  
 This paper is organized as follows: Sect. \ref{sec2} sets up the holographic model for the background and for computing the odd-spin glueball spectra. Sect. \ref{ce1} studies AdS/QCD and CE, applied to odd-spin glueball states. The mass spectra of odd-spin glueballs of higher spin are derived and compared with AdS/QCD predictions. Sect. \ref{sec3} comprises the main results and concluding remarks. 

\section{Holographic Setup and Glueball Spectra}
\label{sec2}

The holographic setup to be considered consists of an Einstein-Hilbert action coupled to a dilaton on the AdS space in five dimensions,
\begin{equation} \label{action}
\!\!\!\!\!\!S \!=\! \dfrac{1}{2\kappa^2}\int  \sqrt{-g}\left(R \!-\! \dfrac{4}{3}g_{\mu\nu}\partial^{\mu}\Phi\partial^{\nu}\Phi \!+\! V(\Phi)\right)\,d^{5}x,
\end{equation}
where $ 2\kappa^2 $ is the gravitational coupling constant, $g$ denotes the determinant of the $g_{\mu\nu}$ metric, $R$ is the (Ricci) scalar curvature, $\Phi$ stands for the dilaton profile and $V(\Phi)$ is the corresponding potential ruling the dilaton field. In this work we will use units such that the AdS radius $L=1$.


For the background, we are going to consider an ansatz in Poincar\'e coordinates, whereas for the dilaton profile the exponential modified dilaton will be taken into account \cite{Fang:2019lmd}, which interpolates between the quadratic dilaton profile, $ \Phi(z) \approx \phi_{\infty}\,z^2 $, in the IR regime ($z\to\infty$), and a quartic dilaton profile, $ \Phi(z) \approx \phi_{\infty}\,z^4 $, in the UV regime ($z\to0$), \blt{with $\phi_{\infty}\sim \Lambda_{QCD}^2$ being the dilaton constant and $ \Lambda_{QCD} $ the energy scale of QCD}. Thus the ansatze we are going to consider are given by: 
\begin{eqnarray}
ds^2 &=& \dfrac{1}{\upzeta(z)^2}\left(dz^2 - dt^2 + dx_i dx^i\right), \label{Metriczeta} \\
\Phi(z) &=& \phi_{\infty}\,z^2\left(1-e^{-\phi_{\infty}\,z^2}\right).\label{expdilaton}
\end{eqnarray}
Therefore, the equations of motion take the form\footnote{For more details we refer the reader to Ref. \cite{MarinhoRodrigues:2020ssq} and references therein.}
\begin{eqnarray}
\frac{\upzeta''(z)}{\upzeta(z)} &=& \frac{4}{9}\Phi'(z)^2,\label{breqn3} \\
V(\Phi) &=& 12\,\upzeta'(z)^2 - \frac{4}{3}\upzeta (z)^2\Phi'(z)^2.\label{Veqn3}
\end{eqnarray}



\begin{figure}[H]
	\centering
	\includegraphics[scale = 0.26]{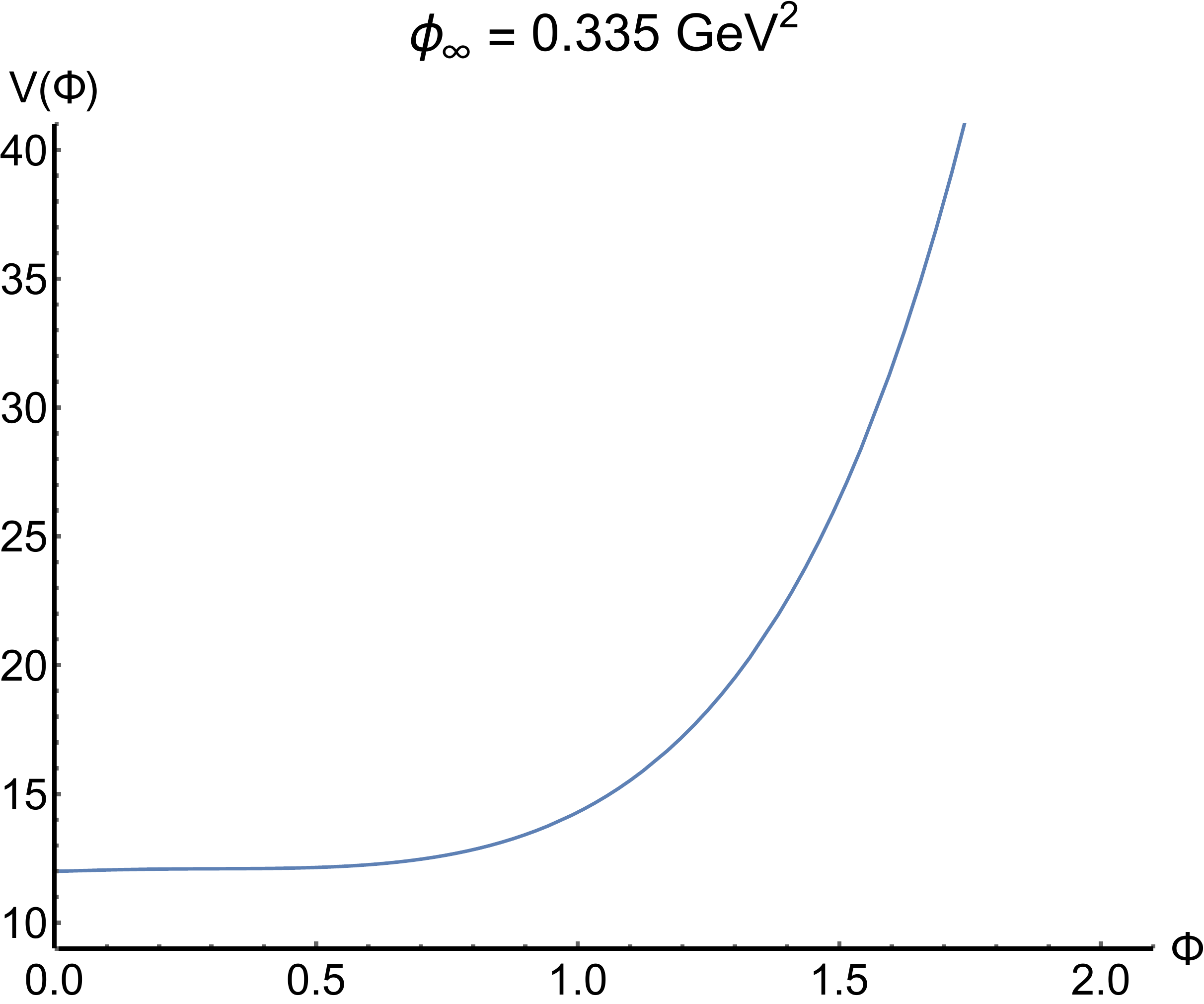}
	\caption{Dilaton potential $V(\Phi)$ for $\phi_{\infty} = 0.335$ GeV$^2$.}
	\label{fig:pot_dilaton}
\end{figure}

\begin{figure}[H]
	\centering
	\includegraphics[scale = 0.26]{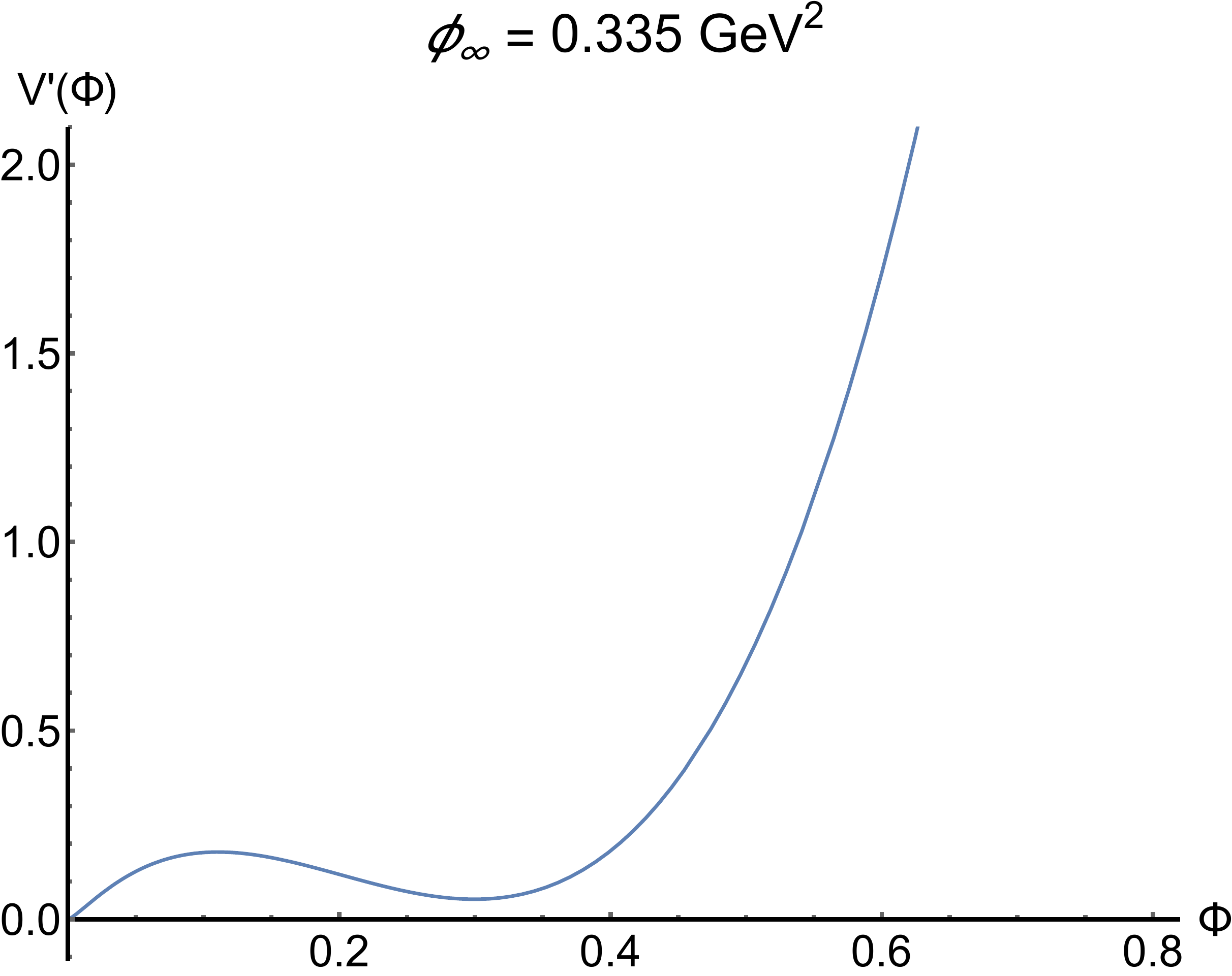}
	\caption{First derivative of the dilaton potential $V(\Phi)$ for $\phi_{\infty} = 0.335$ GeV$^2$.}
	\label{fig:pot_dilatonprime}
\end{figure}

Eqs. (\ref{breqn3}, \ref{Veqn3}) cannot be analytically solved, for the dilaton profile given by \eqref{expdilaton}. Then one needs to solve them numerically, with the boundary condition $\upzeta(z)\approx z$. 
In Figs. \ref{fig:pot_dilaton} and \ref{fig:pot_dilatonprime}, the dilaton potential $V(\Phi)$, given by \eqref{Veqn3}, and its first derivative as a function of $\Phi$, are respectively displayed, for a fixed value of $\phi_{\infty}$, which in this work will be set to $\phi_{\infty} =  0.335$ GeV$^2$. In fact, we have checked that for $\phi_{\infty} \geq 0.335$ GeV$^2$ the dilaton potential is a monotonically increasing function of $\Phi$ \cite{Gursoy:2007cb}.

Having dynamically constructed the background, we turn now to the holographic setup for computing the odd-spin glueball spectra, using the exponential dilaton profile given by Eq. \eqref{expdilaton}.

In the string frame, the glueball action within the dynamical softwall model is given by \blt{$S \!=\!\! \int \sqrt{-g}{\scalebox{1}{$L$}}$}, where 
\begin{equation}\label{acao_soft1}
\blt{\!\!\!\!\!{\scalebox{1}{$L$}}=e^{-\Phi(z)}\left(g_{MN} \partial^M{\mathfrak{G}}\partial^N{\mathfrak{G}} + {\cal M}^2_{5}{\mathfrak{G}}^2\right)},
\end{equation}
where $ {\cal M}_5$ is the mass of the scalar field $ {\mathfrak{G}}(z, x^{\mu}) $, and $\Phi(z)$ is given by Eq. \eqref{expdilaton}.


The field equation coming from \blt{the action computed from} \eqref{acao_soft1} can be put into a Schr\"odinger-like form, by using the following ansatz for the scalar field $ {\mathfrak{G}}(z, x^{\mu})$:
\begin{eqnarray}
{\mathfrak{G}}(z,x^{\mu}) &=& e^{iq^{\mu}x_{\mu}\!+\!{B(z)}/{2}}\psi(z),\label{stat}
\end{eqnarray}
where 
\begin{eqnarray}
B(z) \!=\! 3\,\log{\upzeta(z)} \!+\! \Phi(z).\!\label{stat1}
\end{eqnarray}
\blt{Eq. (\ref{stat}) represents a Bogoliubov transformation of the glueball scalar field followed by a 4-dimensional  Fourier transform, denoted by $\tilde{\mathfrak{G}}(z,q^{\mu})$ \cite{Colangelo:2007pt}. Besides, Eq. (\ref{stat}) consists of a bulk field decomposition, where the $\psi(z)$ is the holographic wave function along the bulk and  the part $e^{iq^{\mu}x_{\mu}}$ stands for a plane wave living on the AdS boundary. Hence, effectively, the scalar field $ {\mathfrak{G}}(z, x^{\mu})$ can be considered as a function of the $z$ coordinate.}  
With this ansatz, one gets the Schr\"odinger-like equation 
\begin{equation}\label{Schrodinger}
-\psi''(z) + V_{\scalebox{.54}{Schr}}(z)\,\psi(z) = (-q^2)\,\psi(z),
\end{equation}
where one identifies $q^2=-m_n^2$, with $m_n$ being the glueball masses, $n = 0, 1,2,\ldots$ represents the radial excitations, whereas  $V_{\scalebox{.54}{Schr}}$ denotes the Schr\"odinger potential,
\begin{equation}\label{SchrPot}
V_{\scalebox{.54}{Schr}}(z) = \left(-\frac{B''}{2}+\frac{B'^2}{4} + {\cal M}_5^2\,\upzeta^{-2}\,e^{\frac{4}{3}\,\Phi(z)}\right).
\end{equation}

From the holographic dictionary and for higher spin fields in AdS, the subsequent mass relation can be regarded,
\begin{equation}\label{massa1}
{\cal M}_5^2 = \Updelta(\Updelta-4) - J + \upgamma(J),
\end{equation} 
where one assumes a contribution coming from the anomalous dimension $\upgamma(J)$ for the glueball operator.
For odd-spin glueballs \cite{Capossoli:2013kb,Capossoli:2016kcr,Capossoli:2016ydo}, the operator that would describe the glueball state $1^{--}$ is given by ${\cal O}_{\Updelta}\equiv { SymTr}\,(\tilde{F}_{\mu\nu}F^2)$, which has conformal dimension $\Updelta=6$. In order to raise the spin, one has to insert the covariant derivatives, to end up with an operator with conformal dimension $\Updelta=6+J$ given by ${\cal O}_{6+J}\equiv SymTr\,(\tilde{F}FD_{\{\mu_1 \cdots}D_{\mu_J\}}\,F)$.
Finally, concerning the anomalous dimensions, we are going to assume a logarithmic dependence on the spin $J$ as 
\begin{eqnarray}
\upgamma(J) &=& \upgamma_0 \log{(1+J)}, \label{AnomalousII}
\end{eqnarray}
where $\upgamma_0$ is a fitting parameter.

With these ingredients, Eq. \eqref{Schrodinger} can be now solved numerically, to obtain the odd-spin glueball mass spectra. \blt{The numerical routine we used is the {\tt{Mathematica}} command {\tt{NDEingensystem}} with Dirichlet boundary conditions.} The results for the mass spectra obtained are displayed in Table  
\ref{massspectra}, for $\phi_{\infty} = 0.335 $ GeV$^2$ fixed, and \blt{five} values of the parameter $\upgamma_0$. In Table \ref{CES5}, we compare our results with the Coulomb gauge QCD model of \cite{LlanesEstrada:2005jf} and the DP Regge model \cite{Szanyi:2019kkn}.

\begin{table}[H]
	\begin{center}\medbreak
		\begin{tabular}{||c||c|c|c|c|c||}
			\hline\hline
			$J^{PC}$ &$1^{--}$&$3^{--}$&$5^{--}$&$7^{--}$&$9^{--}$
			\\\hline\hline
			
			$\blt{\upgamma_0=-20}$~ \,&\,\blt{3.06}\,&\,\blt{4.01}\,&\,\blt{5.80}\,&\,\blt{7.66}\,&\,\blt{8.61} \\\hline
			$\upgamma_0=-24$~ \,&\,2.64\,&\,3.39\,&\,5.29\,&\,7.23\,&\,8.53 \\\hline
			$\upgamma_0=-26$~ \,&\,2.40\,&\,3.03\,&\,5.01\,&\,7.00\,&\,8.49 \\\hline
			$\upgamma_0=-28$~ \,&\,2.13\,&\,2.61\,&\,4.71\,&\,6.77\,&\,8.45 \\\hline
			$\blt{\upgamma_0=-30}$~ \,&\,\blt{1.81}\,&\,\blt{2.09}\,&\,\blt{4.40}\,&\,\blt{6.52}\,&\,\blt{8.40}   \\\hline
			\hline
		\end{tabular}\caption{Mass spectra (GeV) of odd-spin glueballs for \blt{five} values of $\upgamma_0$ and $\phi_{\infty}=0.335$ GeV$^2$.}
		\label{massspectra}
	\end{center}
\end{table}

\begin{table}[H]
	\begin{center}\medbreak
		\begin{tabular}{||cc||c||c||c||c|||c|}
			\hline\hline
			&   $J^{PC}$ & Mass \scriptsize{[this work]} & Mass \scriptsize{[Ref. \cite{LlanesEstrada:2005jf}]} & Mass \scriptsize{[Ref. \cite{Szanyi:2019kkn}]} \\\hline\hline
			\,&\, $1^{--}$ \,&\,  2.40 \,& 3.95  \, &  \\\hline
			\,&\, $3^{--}$ \,&\,  3.03 \,& 4.15 \, & 3.001 \\\hline
			\,&\, $5^{--}$ \,&\,  5.01 \,& 5.05 \, & 4.416 \\\hline
			\,&\, $7^{--}$ \,&\,  7.00 \,& 5.90 \, & 5.498 \\\hline
			\,&\, $9^{--}$ \,&\,  8.49 \,& \,&  \\\hline
			\hline\hline
		\end{tabular}\caption{Mass spectra (GeV) of odd-spin glueballs within the exponential dilaton AdS/QCD model with $\upgamma_0=-26$ and $\phi_{\infty}=0.335$ GeV$^2$ (second column), Coulomb gauge QCD model \cite{LlanesEstrada:2005jf} (third column), and in DP Regge model \cite{Szanyi:2019kkn} (fourth column).}
		\label{CES5}
	\end{center}
\end{table}
Table \ref{CES5} shows the odd-spin glueballs mass spectra for all these models. A bigger gap of $\upgamma_0$ was considered and the results are illustrated in Figs. \ref{figcej} and \ref{figcexm2}. Besides, Table \ref{CEM2} illustrates the new results for the odd-spin glueballs mass spectra.

\blt{Regarding the spectra displayed in Tables \ref{massspectra} and \ref{CES5}, some comments are in order here. The parameter $\phi_{\infty}$ is constrained by the requirement of monotonicity of the dilaton potential, so that we fixed it to the threshold value ($\phi_{\infty} = 0.335 $ GeV$^2$). In fact, we have checked that for $\phi_{\infty} > 0.335 $ GeV$^2$ the mass spectrum increases for any fixed $\upgamma_0$. Therefore, $\upgamma_0$ is the only free parameter one can tune to obtain a spectra compatible with different model calculations. The different values of $\upgamma_0$ presented in Table \ref{massspectra} serve just to illustrate that one  gets a sensible spectra only for $\upgamma_0<0$. In Table \ref{CES5}, for the specific value of $\upgamma_0=-26$, our results are more compatible with those presented in Ref. \cite{Szanyi:2019kkn}. However, if one chooses $\upgamma_0=-20$ (see Table \ref{massspectra}), one can obtain a spectra which matches quite well with lattice QCD. 
 Besides, note that the mass gap, defined as $\delta \equiv m_{J+2}-m_{J}$ ($ J = 1,3,\ldots,9 $), is larger than the other model calculations displayed in Table \ref{CES5}. This is definitely associated with the aforementioned constraint on $\phi_{\infty}\sim\Lambda_{QCD}^2$ due to the dilaton field adopted here. We have checked numerically that for $\phi_{\infty}<0.335$ GeV$^2$ the mass gap $\delta$ gets smaller. However, in this range the dilaton potential $V(\Phi)$ is not a monotonically increasing function of $\Phi$.} 	

\blt{It is worth also noting that in the third column of Table \ref{CES5}, related to the results of \cite{Szanyi:2019kkn}, the glueball state $1^{--}$ is missing. This is so because it is still a matter of debate whether this state belongs or not to the odderon Regge trajectory, in the same way as whether the $0^{++}$ belongs or not to the pomeron Regge trajectory as well. In Ref. \cite{Szanyi:2019kkn} the authors did  consider none of these states in the spectra of even- and odd-spin glueballs. Here in this work it is not our main goal to go beyond with this single discussion.}

\begin{figure}[H]
	\centering
	\includegraphics[scale = 0.26]{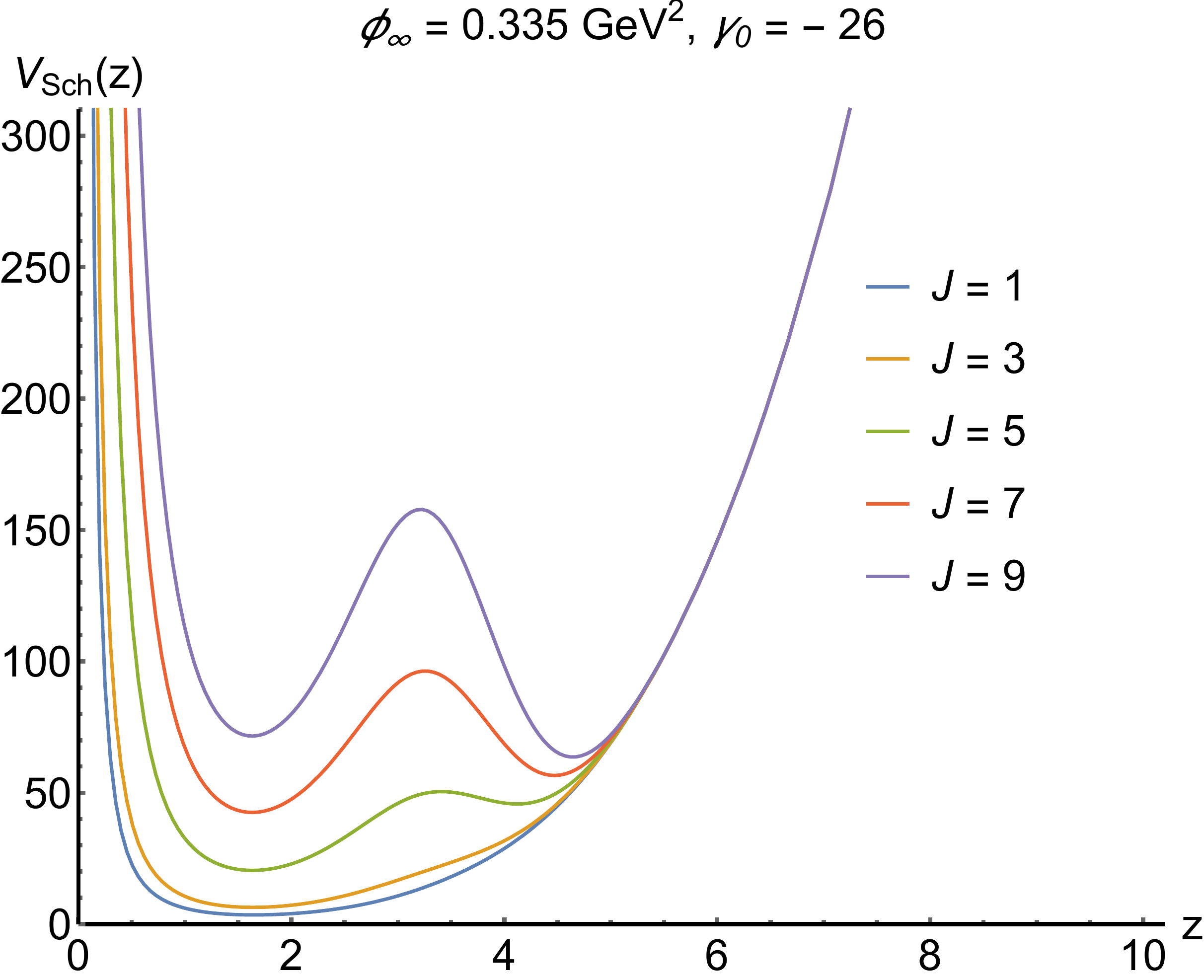}
	\caption{Schr\"odinger potential \eqref{SchrPot} as a function of the Poincar\'e coordinate $z$, for fixed values of $\phi_{\infty}$ and $\upgamma_0$ and for several values of odd-spin $J$, up to $J=9$.}
	\label{fig:pot_dilatonSchr}
\end{figure}

Finally, in Fig. \ref{fig:pot_dilatonSchr}, we show the Schr\"odinger potential \eqref{SchrPot}, for several values of odd-spin spin $J$ up to $J=9$, for fixed values of $\phi_{\infty}$ and $\upgamma_0$.

\section{CE Regge trajectories of odd-spin glueball resonances and mass spectra}\label{ce1}

When all the configurations of a physical system have equal weight, the CE is precisely given by the Boltzmann entropy $S= k_B  \log W$, in the discrete case, 
where $k_B$ denotes the Boltzmann constant and $W$ stands for the number of possible configurations. If the system is split into $n$ states, with respective probabilities $f_n$, the CE reads 
\beq
 S = - k_B  \sum_{n=1}^W f_n \log f_n. 
\eeq
When disorder sets in, equivalently asserting that $f_n = 1/W$,  the standard Boltzmann formula is achieved. Contrarily, when there a single mode configuration with unit probability, the entropy goes to zero. This formulation is the so-called  Gibbs entropy formula, which is analogous to the Shannon information entropy \cite{Bernardini:2016hvx}. \blt{The definition  of the CE is analog  to the Gibbs entropy, defined
of nonequilibrium thermodynamics for a statistical ensemble of microstates.}
For describing a physical system, the energy density $T_{00}({\bf{r}})=\uprho({\bf{r}})$, for ${\bf{r}}\in\mathbb{R}^p$, plays a important role, where $T_{\mu\nu}$ denotes the energy-momentum tensor that encodes the system. The 2-point correlation function ${\scalebox{.9}{$F$}}({\bf{r}})=\int_{\mathbb{R}^{{}^p}} \uprho({\bf{r}}')\uprho({\bf{r}}+{\bf{r}}')\,{\rm d}^p r$  then yields the CE to correspond to the  information entropy proposed by Shannon \cite{Braga:2018fyc}.

The first step in the calculation of the CE is to consider the Fourier transformation,  
\beq\label{fou}
\mathbf\uprho({\bf{k}}) = (2\pi)^{-n/2}\int_{\mathbb{R}^{{}^p}}\mathbf\uprho({\bf{r}})e^{-i{\bf{k}}\cdot {\bf{r}}}\,{\rm d}^px\,.\eeq 
In the CE protocol, the norm of Eq. \eqref{fou} yields the square root of the power spectral density. Therefore, the modal fraction, that is equivalent to the correlations among modes in the system,  reads    
\cite{Gleiser:2012tu}
\begin{eqnarray}
\ringring{\mathbf\uprho}({\bf{k}}) = \frac{\left|\mathbf\uprho({\bf{k}})\right|^{2}}{ \int_{\mathbb{R}^{{}^p}}  |\uprho({\bf{k}})|^{2}{\rm d}^p{k}}.\label{modalf}
\end{eqnarray} 
\blt{The modal fraction stands for the normalized Fourier transform of the 2-point correlation function of the field fluctuations. It then describes the probability distribution for the field fluctuations, having high-power modes more likely to happen than low-power modes. It emulates the probability of a symbol occurrence an alphabet, in Shannon's theory, where the symbol is depicted by a mode in momentum space. } The modal fraction also carries the input of all modes with momentum ${\bf{k}}$ that contribute to the energy density profile. 
Accordingly, the CE for this continuum case is given by the following expression, 
\blt{\begin{eqnarray}
{\rm CE}_\mathbf\uprho= - \int_{\mathbb{R}^{{}^p}}\mathcal{D}({\bf{k}})\, {\rm d}^pk\,,
\label{confige}
\end{eqnarray}
where 
\beq\label{confige1}
\mathcal{D}({\bf{k}})={\mathbf\uprho_{\scalebox{.5}{$\flat$}}}({\bf{k}})\log  {\mathbf\uprho_{\scalebox{.6}{$\flat$}}}({\bf{k}})
\eeq
is the CE density,  
and }
$\uprho_{\scalebox{.6}{$\flat$}}({\bf{k}})=\ringring\uprho({\bf{k}})/\ringring\uprho_{\scalebox{.7}{max}}({\bf{k}})$, for $\ringring{\uprho}_{\scalebox{.7}{max}}$ denoting the supremum among all possible values of the modal fraction. \blt{
The CE is susceptive of  spatial correlations of the physical system at different distances, thus providing an intrinsic informational
measure of spatial coherence in the system. The CE also represents a measure of how compressed is the field, as it can quantify 
the (average) information, in the momentum space, that is needed to portray the field state. Indeed,  any small number of more likely 
modes decreases the CE, also increasing the field compressibility. 
}

To start investigating odd-spin glueballs with the CE, let us consider $p=1$, corresponding to the  $z$ conformal coordinate along the bulk. 
Having the Lagrangian in Eq. \eqref{acao_soft1} for odd-spins, and   
\blt{ \begin{equation}
 \!\!\!\!\!\!\!\!T^{\alpha\beta}\!=\!  \frac{2}{\sqrt{ -g }}\!\! \left[\frac{\partial \left(\sqrt{-g}{{\scalebox{1}{$L$}}}\right)}{\partial{ g_{\alpha\beta}}}\!-\!\frac{\partial}{\partial{ x^\upgamma }}  \frac{\partial{\left(\sqrt{-g} {{\scalebox{1}{$L$}}}\right)}}{\partial \left({\frac{\partial g_{\alpha\beta} }{\partial x^\upgamma}}\right) }
  \right],
  \label{em1}
 \end{equation}}
\!\!\!\!\!\! are the components of the energy-momentum tensor. Hence,   \beq\label{t00}
\uprho(z)= T_{00}(z)=\frac{1}{\upzeta(z)^2}\,\left[\blt{\mathfrak{G}'}^{2}(z) + M_5^2{\mathfrak{G}}^2(z)\right]\,
\eeq
\blt{It is worth to emphasize that Eq. (\ref{stat}) exhibits the glueball bulk scalar field ${\mathfrak{G}}(z,x^{\mu})$ and Eq. (\ref{t00}) regards the energy density representing the odd-spin glueball system in AdS/QCD. 
Also, Eqs. (\ref{stat}, \ref{stat1}) represent the glueball bulk scalar field through a 4-dimensional  Fourier transform with respect to the boundary coordinates yielding  the Schr\"odinger-like equation 
\eqref{Schrodinger} that provides a tower of glueball masses, when  one identifies $q^2=-m_n^2$. Since this equation just depend on the $z$ coordinate, to compute the CE one integrates along the energy scale, that corresponds to the bulk extra dimension. Hence the 4-dimensional coordinates do not effectively appear here, being the scalar field  described by ${\mathfrak{G}}(z)$.}

The CE underlying the odd-spin glueballs can be immediately calculated by using Eqs. (\ref{fou} -- \ref{confige}), for any odd  value  of  $J$.
Moreover, a first type of configurational-entropic (CE) Regge trajectory, that associates the CE of odd-spin glueballs to their $J$ spin, can be engendered. Table \ref{CEJ} compiles the resulting data, \blt{numerically computed}. Hereon, for conciseness, we denote \blt{$\upgamma_{0_0}=-22$, $\upgamma_{0_1}=-24$, $\upgamma_{0_2}\equiv -26$, $\upgamma_{0_3}\equiv -28$, and $\upgamma_{0_4}\equiv -30$}.
\begin{table}[H]
\begin{center}\medbreak
\begin{tabular}{||c||c|c|c|c|c|c|c|c||}
\hline\hline
   $J$~& 1&3&5&7&9&11*&13*&15*
    \\\hline\hline
   \blt{CE$_{{\scalebox{.75}{$\upgamma_{0_0}$}}}$}~ &3.08&5.01&13.91&31.89&62.10&118.08&187.35&259.36 \\\hline
     \blt{CE$_{{\scalebox{.75}{$\upgamma_{0_1}$}}}$}~ &3.22&5.34&14.59&33.91&65.40&111.83&175.64&275.73 \\\hline
     CE$_{{\scalebox{.75}{$\upgamma_{0_2}$}}}$~ &3.50&5.64&15.06&35.01&68.29&118.08&187.35&279.15 \\\hline
     CE$_{{\scalebox{.75}{$\upgamma_{0_3}$}}}$&3.93&6.54&18.71&39.65&74.24&122.85&188.10&271.96  \\\hline
      \blt{CE$_{{\scalebox{.75}{$\upgamma_{0_4}$}}}$}~ &4.22&7.03&20.53&42.81&77.25&122.99&181.32&252.86 \\\hline
    \hline
    \end{tabular}\caption{CE of odd-spin glueballs as a function of  $J$. The CE of odd-spin glueballs states with $J=11, 13, 15$, indicated with an asterisk, are extrapolated from the CE Regge trajectories (\ref{itp1exp22} -- \ref{itp1exp30}).}
\label{CEJ}
\end{center}
\end{table}
In Table \ref{CEJ}, the values of the CE for $J>9$ can be  extrapolated 
by interpolation of the previous values of the CE up to $J=9$. It yields, respectively, 
the  CE Regge trajectories,  
 \begin{eqnarray}
  \label{itp1exp22}
\!\!\!\!\!\!\! \blt{{\rm CE}_{\scalebox{.71}{$\upgamma_{0_0}$}}(J)} \!&\!=\!&\blt{\! 0.0547\,J^3\!+\!0.3502\,J^2\!-\!1.1112\,J}\nonumber\\&&\qquad\qquad\qquad\qquad\qquad\blt{+3.7713}, \\
  \label{itp1exp24}
\!\!\!\!\!\!\! \blt{{\rm CE}_{\scalebox{.71}{$\upgamma_{0_1}$}}(J)} \!&\!=\!&\! \blt{0.0851\,J^3\!+\!0.1246\,J^2\!-\!0.5460\,J}\nonumber\\&&\qquad\qquad\qquad\qquad\qquad\blt{+3.5561}, \\
 \label{itp1exp}
\!\!\!\!\!\!\! {\rm CE}_{\scalebox{.71}{$\upgamma_{0_2}$}}(J) \!&\!=\!&\! 0.0631\,J^3\!+\!0.3535\,J^2\!-\!1.1813\,J\nonumber\\&&\qquad\qquad\qquad\qquad\qquad+4.2724, \\
\!\!\!\!\!\!\! {\rm CE}_{\scalebox{.71}{$\upgamma_{0_3}$}}(J) \!&\!=\!&\! 0.1149\,J^3\!+\!0.1592\, J^2\!-\!0.8251\,J\nonumber\\&&\qquad\qquad\qquad\qquad\qquad+4.4859, \label{itp1exp2}\\
\!\!\!\!\!\!\! \blt{{\rm CE}_{\scalebox{.71}{$\upgamma_{0_4}$}}(J)} \!&\!=\!&\! \blt{0.1290\,J^3\!+\!0.1744\, J^2\!-\!0.9708\,J}\nonumber\\&&\qquad\qquad\qquad\qquad\qquad\blt{+4.8873}. \label{itp1exp30}
      \end{eqnarray}     \noindent   
Table \ref{CEJ} and the CE Regge trajectories (\ref{itp1exp22} -- \ref{itp1exp30}) are illustrated in Fig. \ref{figcej}. 
\begin{figure}[H]
\centering
\includegraphics[width=7cm]{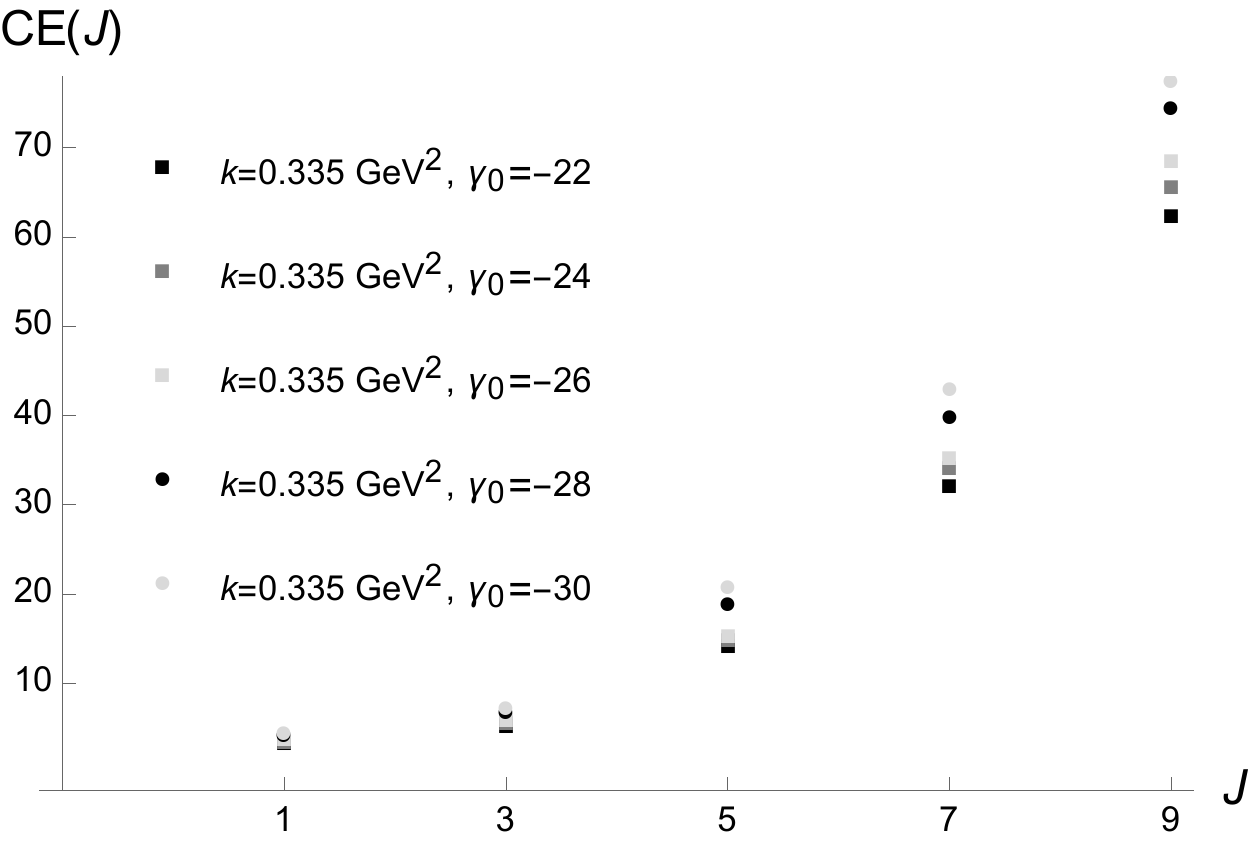}
\caption{CE of odd-spin glueballs  as a function of $J$, for \blt{five} values of $\upgamma_0$, and the respective CE Regge trajectories.}
\label{figcej}
\end{figure}
\noindent

Now, to derive the odd-spin glueball mass spectra, we 
can take data in Table \ref{CEJ} and compute the CE as a function of the odd-spin glueball mass spectra for $J\leq 9$. Therefore, the mass 
spectra of odd-spin glueballs with $J=11, 13, 15,
\ldots$ can be obtained by interpolation methods. 
These second kinds of CE Regge trajectories are depicted is Fig. \ref{figcexm2}, whose interpolation formul\ae\, are shown in Eqs. (\ref{itq1exp22} -- \ref{itq1exp30}).
\begin{figure}[H]
\centering
\includegraphics[width=7.5cm]{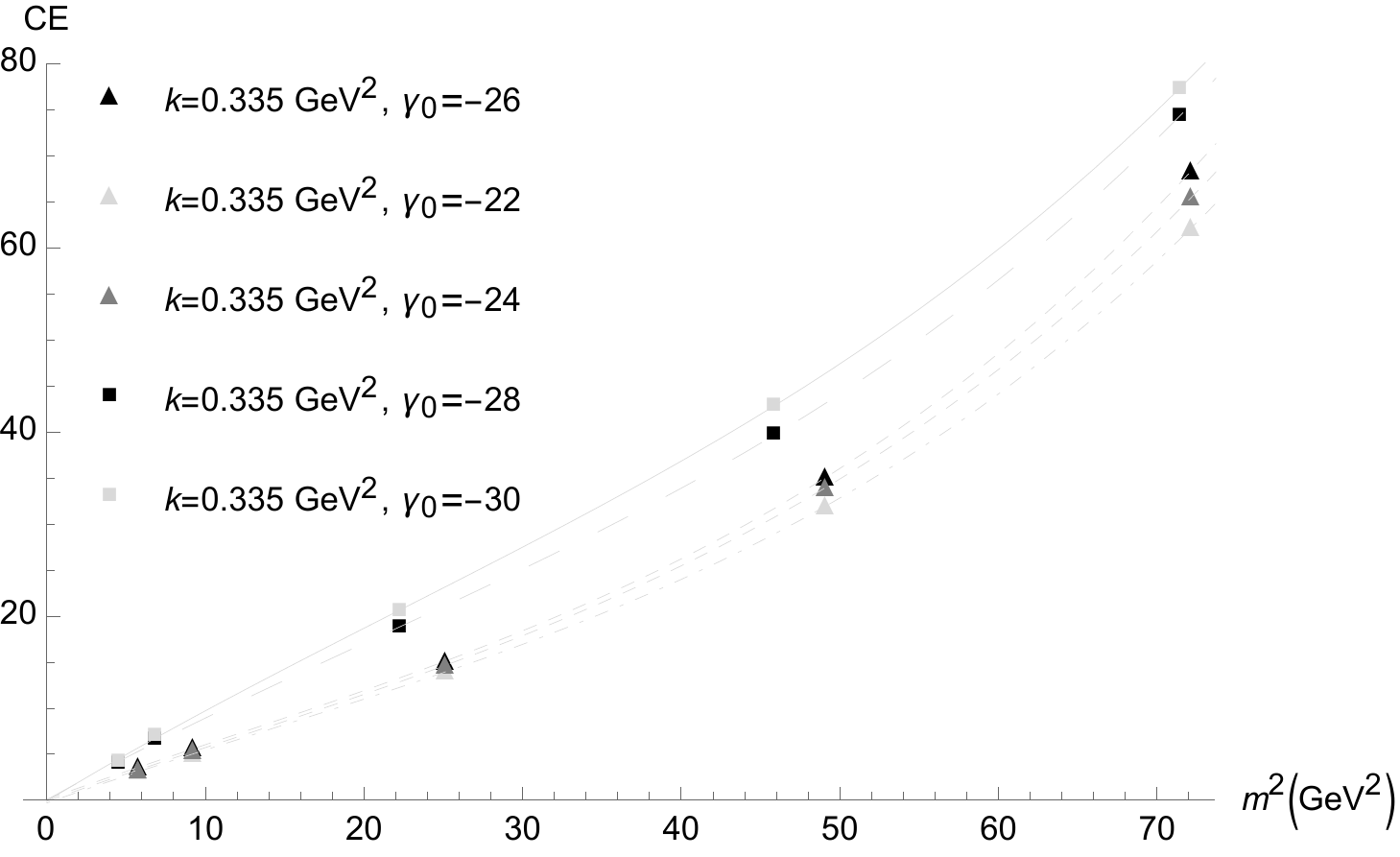}\medbreak
\caption{CE of odd-spin glueballs, with respect to the mass spectra, and their respective CE Regge trajectories. }
\label{figcexm2}
\end{figure} \noindent Odd-spin glueball masses, for any value of $J$ can be obtained when one extrapolates the CE Regge trajectories,  
 \begin{eqnarray}
  \label{itq1exp22}
\!\!\!\!\!\!\! \blt{{\rm CE}_{\scalebox{.71}{$\upgamma_{0_0}$}}(m)} \!&\!=\!&\! \blt{0.0001\,m^6\!-\!0.0046\,m^4\!+\!0.6112\,m^2}\nonumber\\&&\qquad\qquad\qquad\qquad\qquad\blt{-0.3082}, \\
  \label{itq1exp24}
\!\!\!\!\!\!\! \blt{{\rm CE}_{\scalebox{.71}{$\upgamma_{0_1}$}}(m)} \!&\!=\!&\! \blt{0.0001\,m^6\!-\!0.0033\,m^4\!+\!0.6034\,m^2}\nonumber\\&&\qquad\qquad\qquad\qquad\qquad\blt{-0.1018}, \\
 \label{itq1exp}
\!\!\!\!\!\!\! {\rm CE}_{\scalebox{.71}{$\upgamma_{0_2}$}}(m) \!&\!=\!&\! 0.0001\,m^6\!-\!0.0037\,m^4\!+\!0.6184\,m^2\nonumber\\&&\qquad\qquad\qquad\qquad\qquad+0.0999, \\
\!\!\!\!\!\!\! {\rm CE}_{\scalebox{.71}{$\upgamma_{0_3}$}}(m) \!&\!=\!&\! 0.0001\,m^6\!-\!0.0064\, m^4\!+\!0.9509\,m^2\nonumber\\&&\qquad\qquad\qquad\qquad\qquad+0.0390, \label{itq1exp2}\\
\!\!\!\!\!\!\! \blt{{\rm CE}_{\scalebox{.71}{$\upgamma_{0_4}$}}(m)} \!&\!=\!&\! \blt{0.0001\,m^6\!-\!0.0076\, m^4\!+\!1.0463\,m^2}\nonumber\\&&\qquad\qquad\qquad\qquad\qquad\blt{-0.1119}, \label{itq1exp30}
      \end{eqnarray} 
constructed upon interpolating the discrete points in Fig. \ref{figcexm2}, with accuracy within 0.1\%.
For $J=11$, Eq. (\ref{itp1exp22}) yields 
${\rm CE}_{\mathbb{O}_{11}} = 118.08$. When one substitutes it into (\ref{itq1exp}) yields $
m_{\mathbb{O}_{11}}=9.65\; {\rm GeV}$, for the $J=11$  $\mathbb{O}_{11}$ odd-spin glueball mass. In a similar way, this procedure  can be accomplished for higher  odd-spin glueballs, for the \blt{five} values of $\upgamma_0$, respectively using the pairs of equations (\ref{itp1exp22} -- \ref{itq1exp30}). The respective mass spectra are illustrated in Table \ref{CEM2}. \blt{Let us remember that we choose $\phi_{\infty} =  0.335$ GeV$^2$, together with the values of $\upgamma_0$ that are in the range $-30\lesssim\upgamma_0\lesssim -22$ that better describes  the odd-spin glueballs mass spectra in AdS/QCD dynamical model.}
\begin{table}[H]
\begin{center}\medbreak
\begin{tabular}{||c||c|c|c||}
\hline\hline
   $J$~&11&13&15
    \\\hline\hline
     \blt{$m_{\scalebox{.71}{$\upgamma_{0_0}$}}$}~ \,&\,9.84\,&\,10.80\,&\,11.49 \\\hline
     \blt{$m_{\scalebox{.71}{$\upgamma_{0_1}$}}$}~ \,&\,9.65\,&\,10.71\,&\,11.60 \\\hline
    $m_{\scalebox{.71}{$\upgamma_{0_2}$}}$\,&\,9.64\,&\,10.61\,&\,11.44  \\\hline
     $m_{\scalebox{.71}{$\upgamma_{0_3}$}}$\,&\,9.65\,&\,10.62\,&\,11.44  \\\hline
         \blt{ $m_{\scalebox{.71}{$\upgamma_{0_4}$}}$}\,&\,9.58\,&\,10.57\,&\,11.34  \\\hline
    \hline
    \end{tabular}\caption{Mass spectra (GeV) of odd-spin glueballs as a function of $J$.  }
\label{CEM2}
\end{center}
\end{table}
\noindent \blt{ The results derived by the previous CE-based analysis, and illustrated in Table \ref{CEM2}, have a precise profile. For fixed values of $J=11,13$, the higher the value of $\upgamma_0$, the lower the odd-spin glueball mass is. However, for $J=15$, the mass is observed to have a maximum at $\upgamma_0=-24$.}

It is worth to mention that this method holds for any odd value of $J$. However we restrict ourselves to $J\leq 15$, since 
higher values of $J$ will be very unlikely to be detected even in future experiments.   One can notice from Table \ref{CEM2} that the \blt{five} different values of the anomalous dimension parameter, $\upgamma_0$, yield almost the same mass spectra. 

One can still compare the odd-spin glueballs mass spectra in Table \ref{CEM2}, to the mass spectra yielded by AdS/QCD in Table \ref{CES5}, \blt{for $\upgamma_0=-26$}. Denoting by $\Updelta_J$ the difference between the mass spectra of odd-spin glueballs, $\Updelta_{11}=4.4\%$ and $\Updelta_{13}=5.9\%$, whereas $\Updelta_{15}=22.1\%$.  The mass spectra in Table \ref{CEM2} regards odd-spin glueballs that are more configurationally stable
than their AdS/QCD counterparts. \blt{For the other values of $\upgamma_0$ heretofore used, similar values of the respective $\Updelta_J$ are obtained.}

\blt{Table \ref{CEJ} illustrates a feature that is worth to emphasize. For every fixed $J$ in Table \ref{CEJ}, the lower the value of $\upgamma_0$, the higher the configurational stability is, for odd-spin glueball states. In fact, for any fixed value of $J$, the higher CE occurs at $\upgamma_0=-30$. Besides, for every fixed value of $\upgamma_0$ in the range studied, the higher the value of $J$, the higher the CE is.  Both profiles indicate two recurrent behaviors. The first one consists of the fact that for any fixed value of $J$, in the range $-30\lesssim\upgamma_0\lesssim -22$, 
higher $J$ values are more unstable, from the CE point of view. Thus, the odd-spin glueballs corresponding to lower values of $\upgamma_0$ prevail, being eventually more likely to be detected. The second feature reveals that for any fixed value of $\upgamma_0$ in the range studied, higher spin states are also more unstable, from the CE point of view, as their CE monotonically increase according to the scale laws (\ref{itp1exp22} -- \ref{itp1exp30}). Hence, the more stable odd-spin glueballs states have mass $m_{11}=9.84$ GeV, $m_{13}=10.80$ GeV, and $m_{15}=11.49$ GeV. }

\blt{\subsection{Mass spectra by lower spin states interpolation}}
\label{subsee}

\blt{The part of Sect. \ref{ce1} presented heretofore predicts the odd-spin glueballs mass spectra for high spin states. Due to lack of robust experimental data regarding odd-spin glueballs, it is also reasonable to consider just the  $J = 1,3,5$ states in AdS/QCD and extrapolate  higher odd-spin glueballs for $J>5$, using numerical interpolation. The odd-spin glueball mass spectra, for the same values of $\upgamma_0$ already chosen, are presented and discussed. We emphasize that when trying to adjust between only two points, $J=1$ and $J=3$, a straight line matching these points would make a poor interpolation. Therefore, a three point-interpolation ($J = 1,3,5$) is the minimally feasible one that can generate a CE-based approach to describe the odd-spin glueball mass spectra.}

\blt{\begin{table}[H]
\begin{center}\medbreak
\begin{tabular}{||c||c|c|c|c|c|c|c|c||}
\hline\hline
   $J$~& 1&3&5&7*&9*&11*&13*&15*
    \\\hline\hline
   \blt{CE$_{{\scalebox{.75}{$\upgamma_{0_0}$}}}$} &3.08&5.01&13.91&38.37&90.54&186.11&344.32&587.97 \\\hline
     \blt{CE$_{{\scalebox{.75}{$\upgamma_{0_1}$}}}$} &3.22&5.34&14.59&39.66&92.81&189.90&350.35&597.14 \\\hline
     CE$_{{\scalebox{.75}{$\upgamma_{0_2}$}}}$ &3.50&5.64&15.06&40.63&94.95&194.25&358.41&611.02 \\\hline
     CE$_{{\scalebox{.75}{$\upgamma_{0_3}$}}}$~&3.93&6.54&18.71&52.18&123.62&254.50&471.18&804.88  \\\hline
      \blt{CE$_{{\scalebox{.75}{$\upgamma_{0_4}$}}}$} &4.22&7.03&20.53&57.96&138.05&285.00&528.49&903.66 \\\hline
    \hline
    \end{tabular}\caption{\blt{CE of odd-spin glueballs as a function of  $J$. The CE of odd-spin glueballs states with $J=7,9,11, 13, 15$, indicated with an asterisk, are extrapolated from the CE Regge trajectories (\ref{1itp1exp22} -- \ref{1itp1exp30}).}}
\label{CEJ1}
\end{center}
\end{table}}

\blt{Also emulating Table \ref{CEJ}, the values of the CE for $J>5$ can be  extrapolated 
by interpolation of the previous values of the CE up to $J=5$. This procedure yields, respectively, 
the  CE Regge trajectories,}  
\blt{ \begin{eqnarray}
  \label{1itp1exp22}
\!\!\!\!\!\!\! \blt{{\rm CE}_{\scalebox{.71}{$\upgamma_{0_0}$}}}(J) \!&\!=\!&\! 0.0092\,J^4\!+\!0.0311\,J^3\!+\!0.05497\,J^2\nonumber\\&&\qquad\qquad\qquad-0.0292\,J+3.01393, \\
  \label{1itp1exp24}
\!\!\!\!\!\!\! \blt{{\rm CE}_{\scalebox{.71}{$\upgamma_{0_1}$}}(J)} \!&\!=\!&\! 0.0093\,J^4\!+\!0.0320\,J^3\!+\!0.0628\,J^2\nonumber\\&&\qquad\qquad\qquad+0.01973\,J+ 3.09606
 \label{1itp1exp}\\
\!\!\!\!\!\!\! {\rm CE}_{\scalebox{.71}{$\upgamma_{0_2}$}}(J) \!&\!=\!&\! 0.0095\,J^4\!+\!0.0325\,J^3\!+\!0.06197\,J^2\nonumber\\&&\qquad\qquad\qquad+0.01728\,J+ 3.3796, \\
\!\!\!\!\!\!\! {\rm CE}_{\scalebox{.71}{$\upgamma_{0_3}$}}(J) \!&\!=\!&\! 0.0126\,J^4\!+\!0.0426\,J^3\!+\!0.0752\,J^2\nonumber\\&&\qquad\qquad\qquad-0.0563\,J+ 3.8606, \label{itp1exp2}\\
\!\!\!\!\!\!\! \blt{{\rm CE}_{\scalebox{.71}{$\upgamma_{0_4}$}}(J)} \!&\!=\!&\! 0.0142\,J^4+0.0476\,J^3\!+\!0.0797\, J^2\nonumber\\&&\qquad\qquad\qquad-0.1044\,J+4.1827. \label{1itp1exp30}
      \end{eqnarray}}     \noindent   
\blt{Table \ref{CEJ1} and the CE Regge trajectories (\ref{1itp1exp22} -- \ref{1itp1exp30}) are illustrated in Fig. \ref{figcej}. }
\begin{figure}[H]
\centering
\includegraphics[width=7.8cm]{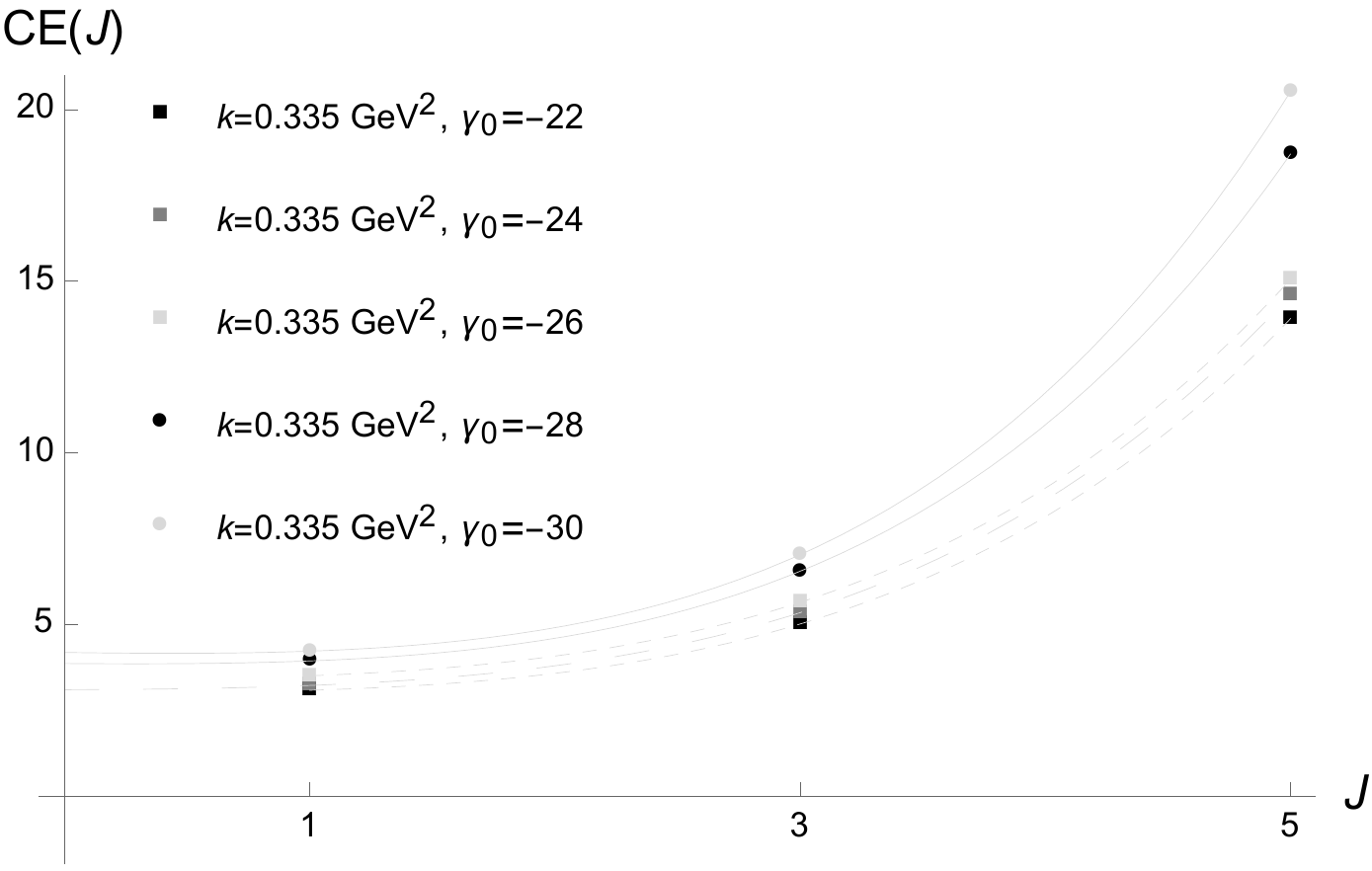}
\caption{\blt{CE of odd-spin glueballs  as a function of $J$, for {four} values of $\upgamma_0$, and the respective CE Regge trajectories.}}
\label{figcej}
\end{figure}
\noindent 
\blt{Based on Table \ref{CEJ}, for $J = 1,3,5$, the mass spectra of odd-spin glueballs with $J\geq 5$ will be thus derived. 
These second kinds of CE Regge trajectories are depicted is Fig. \ref{figcexm2}, whose interpolation formul\ae\, are shown in Eqs. (\ref{1itq1exp22} -- \ref{1itq1exp30}).}
\begin{figure}[H]
\centering
\includegraphics[width=8.2cm]{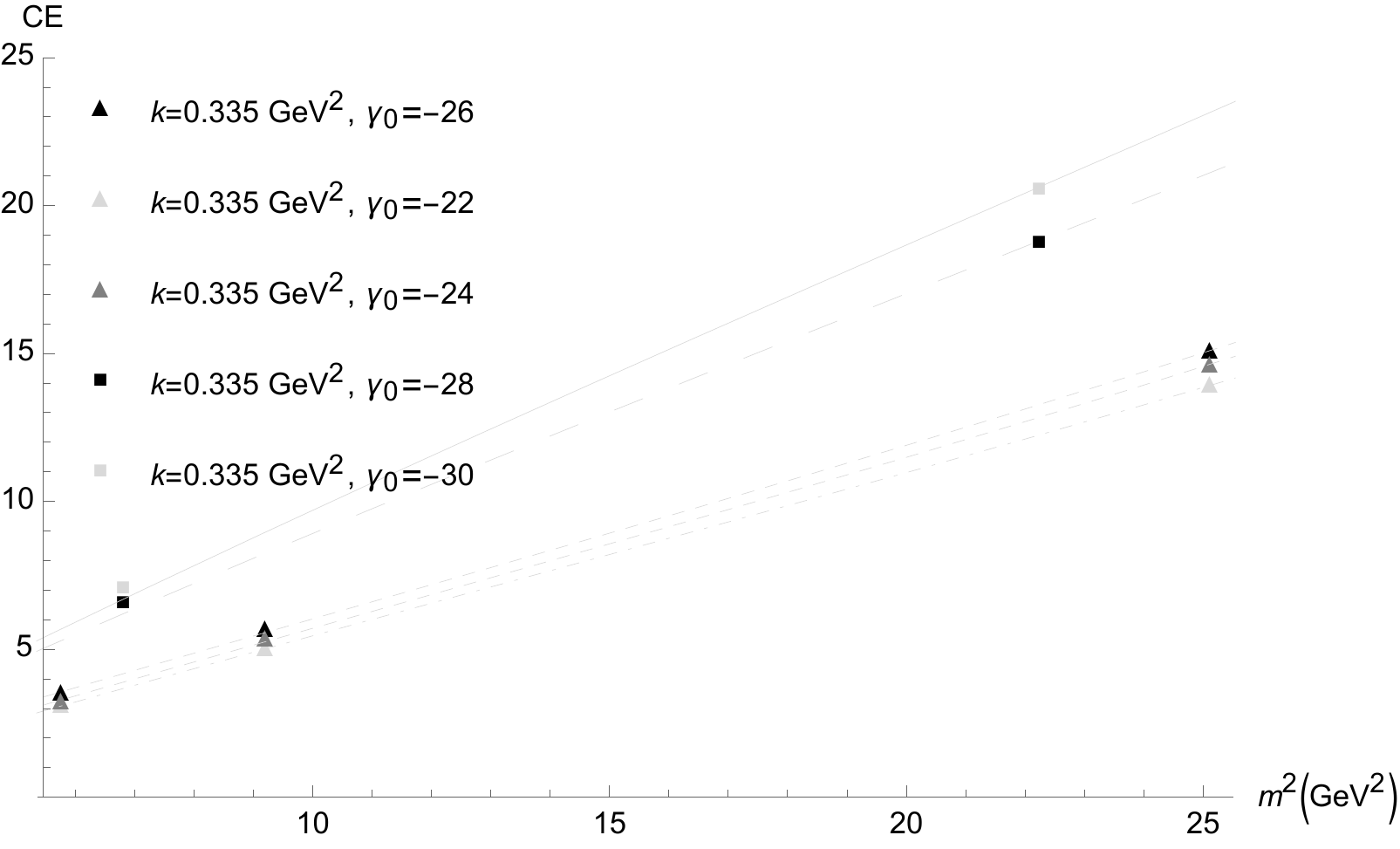}\medbreak
\caption{\blt{CE of odd-spin glueballs, with respect to the mass spectra, and their respective CE Regge trajectories, in the three-point ($J=1,3,5$) interpolation.} }
\label{figcexm2}
\end{figure} \noindent \blt{Odd-spin glueball masses, for any value of $J$ can be obtained when one extrapolates the CE Regge trajectories,  
 \begin{eqnarray}
  \label{1itq1exp22}
\blt{{\rm CE}_{\scalebox{.71}{$\upgamma_{0_0}$}}(m)} \!&\!=\!&\! -1.7639\!\times\! 10^{-4}\,m^6\nonumber\\&&\quad0.0069\,m^4\!\!+\!0.4891\,m^2\!+0.0661, \\
  \label{1itq1exp24}
\blt{{\rm CE}_{\scalebox{.71}{$\upgamma_{0_1}$}}(m)} \!&\!=\!&\! -2.2913\!\times\! 10^{-4}\,m^6\nonumber\\&&\quad+0.0072\,m^4\!\!+\!0.5477\,m^2\!-0.1340, \\
\label{1itq1exp}
\blt{{\rm CE}_{\scalebox{.71}{$\upgamma_{0_2}$}}(m)} \!&\!=\!&\! -2.2573\,\!\times\! 10^{-4}\,m^6\nonumber\\&&\quad+0.0073\,m^4\!\!+\!0.5521\,m^2\!+\!0.1184, \\
\blt{{\rm CE}_{\scalebox{.71}{$\upgamma_{0_3}$}}(m)} \!&\!=\!&\! -8.7253\!\times\! 10^{-4}\,m^6\!\nonumber\\&&\quad+0.0091\,m^4\!+\!1.1263\,m^2\!-\!1.2776, \label{1itq1exp2}\\
\blt{{\rm CE}_{\scalebox{.71}{$\upgamma_{0_4}$}}(m)} \!&\!=\!&\! -9.0393\!\times\! 10^{-4}\,m^6\!\nonumber\\&&\quad+0.0102\,m^4\!+\!1.2042\,m^2\!-\!1.3644. \label{1itq1exp30}
      \end{eqnarray} }   
\blt{Such trajectories are constructed when one  interpolates the discrete points in Fig. \ref{figcexm2}, with accuracy within 0.16\%.
Higher  (odd-)spin glueballs, for $J>5$, for the five values of $\upgamma_0$, respectively using the pairs of equations (\ref{1itp1exp22} -- \ref{1itq1exp30}). The derived mass spectra are illustrated in Table \ref{CEM21}. }
\blt{\begin{table}[ht]
\begin{center}\medbreak
\begin{tabular}{||c||c|c|c|c|c||}
\hline\hline
   $J$~&7&9&11&13&15
    \\\hline\hline
     {$m_{\scalebox{.71}{$\upgamma_{0_0}$}}$}~ &7.87&8.87&9.91&10.95&11.96 \\\hline
    {$m_{\scalebox{.71}{$\upgamma_{0_1}$}}$}~ &7.67&8.61&9.59&10.57&11.55 \\\hline
    $m_{\scalebox{.71}{$\upgamma_{0_2}$}}$&7.70&8.65&9.64&10.63&11.60  \\\hline
     $m_{\scalebox{.71}{$\upgamma_{0_3}$}}$&6.79&7.53&8.32&9.12&9.91  \\\hline
        { $m_{\scalebox{.71}{$\upgamma_{0_4}$}}$}&6.85&7.60&8.41&9.22&10.12  \\\hline
    \hline
    \end{tabular}\caption{\blt{Mass spectra (GeV) of odd-spin glueballs as a function of $J$.}  }
\label{CEM21}
\end{center}
\end{table}}

\blt{The results for the three-point interpolation method are qualitatively similar to those ones obtained at the beginning of Sect. \ref{ce1}. As illustrated in Table  \ref{CEJ1}, higher values of $J$ are configurationally more unstable. Analogously to the previous results of Sect. \ref{ce1}, odd-spin glueballs related to lower values of $\upgamma_0$  are also more dominant. Now, taking any fixed value of $\upgamma_0$ in the range studied, higher spin  odd-spin glueballs are less stable from the CE point of view. Then the more stable odd-spin glueballs states have mass respectively given by $m_{7}=7.87$ GeV, $m_{9}=8.87$ GeV,  $m_{11}=9.91$ GeV, $m_{13}=10.95$ GeV, and $m_{15}=11.96$ GeV. The three-point interpolation results of the mass spectra, for the more stable states, differs from the five-point interpolation in $0.7\%$, for $J=11$; in $1.4\%$, for $J=13$, and in $4.1\%$, for $J=15$. Although the masses have  relatively near values in both interpolations, the three-point interpolation in Subsect. \ref{subsee} constitutes a more realistic method: deriving also the mass spectra for $J=7$ and $J=9$ constitutes the prediction of states that are more likely to be detected or observed in experiments. One can again compare the odd-spin glueballs mass spectra in Table \ref{CEM21}, to the mass spectra yielded by AdS/QCD in Table \ref{CES5}. One has $\Updelta_{7}=14.8\%$, $\Updelta_{9}=17.7\%$, $\Updelta_{11}=19.1\%$ and $\Updelta_{13}=20\%$, whereas $\Updelta_{15}=20.6\%$. }

\blt{Now, we would like to discuss an important issue. Lightest glueballs are mainly two-gluon states. The more gluons constitute a glueball, the heavier the glueball is. The scalar glueball $0^{++}$, has the same quantum number as the scalar meson ($\bar{q}q$) and tetraquark ($\bar{q}q\bar{q}q$) states, and the complexity of deriving pure glueball states underlies in the fact that gluon bound states might mix with $\bar{q}q$ and $\bar{q}q\bar{q}q$ states since, in addition to the quantum numbers, their energy scales are very similar (for more details on the glueball-meson mixing we refer the reader to \cite{Mathieu:2008me, Vento:2015yja} and references therein). This process also includes odd-spin glueballs and higher $J$ states \cite{deMelloKoch:1998vqw}. However, the dynamical AdS/QCD model here discussed does not take into account the odd-spin glueballs intrinsic constitution and the possible mixing with meson states. More recently, an interesting approach, also based on AdS/QCD, has been proposed by the authors of \cite{Rinaldi:2018yhf,Rinaldi:2020ssz,Rinaldi:2017wdn}, in which they gave an interpretation for the mixing mechanism by assuming a light-front quantum-mechanical holographic description. They have shown that the overlap probability of heavy glueballs to heavy mesons is small. Therefore, one could expect a scenario of little or no mixing at high energies, more specifically, at energies above $2$ GeV \cite{Rinaldi:2018yhf,Rinaldi:2020ssz}. Thus, in this kinematical regime one could expect to obtain states with mostly gluonic valence structure, i.e., pure glueball states. In this sense, the higher odd-spin glueball states scrutinized in this work are expected to be of relevant phenomenological interest.}  

\blt{\section{Concluding remarks and perspectives}}\label{sec3}

\blt{In this work we have considered a CE-based method to predict the masses of higher odd-spin glueball states based on a dynamical AdS/QCD model with a modified exponential dilaton model having logarithmic anomalous dimension corrections. The conciseness of the CE-based analysis presented here circumvents intricacies that are inherent when studying phenomenological methods in AdS/QCD. The mass spectra obtained from both AdS/QCD model and from CE analysis are compatible with a recent result from Regge theory as well as lattice QCD.}

\blt{According to Table \ref{CEJ}, for fixed values of $J$, the lower the value of $\upgamma_0$, the higher the configurational stability is, underlying the studied odd-spin glueballs. In addition, for fixed values of $\upgamma_0$, the higher the value of $J$, the higher the CE is.  It yields the conclusion that higher values of $J$ are configurationally more unstable, being lower values of $\upgamma_0$ more prevalent. This feature is recurrent in both the three- or the five-point interpolation methods in Sect. \ref{ce1}.}

\blt{It is worth to emphasize that PDG does not provide any data for odd-spin glueballs, up to our knowledge. Hence, the main relevance of this work consists of engendering an applicable database for future experiments and probing further aspects in order to better understand the physics of glueballs. Furthermore, as already shown in the literature, the higher spin glueball states, if they exist in Nature, are expected to be constituted with mostly gluonic valence structure and thus being pure glueball states. Therefore, computing their mass spectra has a relevant phenomenological interest.}

\blt{Finally, an interesting perspective would be to use a light-front holographic approach combined with CE-based methods in order to deeply investigate the glueball-meson mixing in order to see whether the CE brings new insights into this problem. We will leave this analysis for future research.}

--------------------------------------------------------------------


\medbreak
\paragraph*{Acknowledgments:}  DMR is supported by the National Council for Scientific and Technological Development -- CNPq (Brazil) under Grant No. 152447/2019-9. RdR is grateful to FAPESP (Grant No. 2017/18897-8) and to CNPq (Grants No. 303390/2019-0, No. 406134/2018-9), for partial financial support. 

\end{document}